\documentclass[preprint, showpacs,preprintnumbers, amsmath,amssymb, aps, prb,]{revtex4-1}

\usepackage{graphicx}
\usepackage{dcolumn}
\usepackage{bm}
\usepackage{hyperref}
\usepackage[cp1250]{inputenc}
\usepackage[usenames,dvipsnames]{color}

\newcommand{\tr}{\textcolor{Red}}
\renewcommand{\tr}{}
\newcommand{\trr}{\textcolor{Red}}
\renewcommand{\trr}{}

\newcommand{\bk}{\mathbf{k}}
\newcommand{\bq}{\mathbf{q}}
\newcommand{\bQ}{\mathbf{Q}}
\newcommand{\bfr}{\mathbf{r}}

\newcommand{\ua}{\uparrow}
\newcommand{\da}{\downarrow}

\newcommand{\mb}{\mathbf}

\newcommand{\eq}{\begin{equation}}
\newcommand{\eqx}{\end{equation}}
\newcommand{\eqn}{\begin{eqnarray}}
\newcommand{\eqnx}{\end{eqnarray}}

\newcommand{\s}{\sigma}

\begin{document}

\preprint{APS/123-QED}

\title{Conductance spectroscopy of a correlated superconductor in a magnetic field in the Pauli limit: Evidence for strong correlations}

\author{Jan Kaczmarczyk}
\email{jan.kaczmarczyk@uj.edu.pl}

\author{Mariusz Sadzikowski}
\email{sadzikowski@th.if.uj.edu.pl}
\affiliation{
Marian Smoluchowski Institute of Physics, Jagiellonian University, \linebreak Reymonta 4, 30-059 Kraków, Poland
}

\author{Jozef Spałek}
 \email{ufspalek@if.uj.edu.pl}
\affiliation{
Marian Smoluchowski Institute of Physics, Jagiellonian University, \linebreak Reymonta 4, 30-059 Kraków, Poland \\
Faculty of Physics and Applied Computer Science, AGH University of Science and Technology, Reymonta 19, 30-059 Kraków, Poland
}

\date{\today}

\begin{abstract}
We study conductance spectroscopy of a two-dimensional junction between a normal metal and a \textit{strongly-correlated} superconductor in an applied magnetic field in the Pauli limit. Depending on the field strength the superconductor is either in the Bardeen-Cooper-Schrieffer (BCS), or in the Fulde-Ferrell-Larkin-Ovchinnikov (FFLO) state of the Fulde-Ferrell (FF) type. The strong correlations are accounted for by means of the Gutzwiller method what leads naturally to the emergence of the spin-dependent masses (SDM) of quasiparticles when the system is spin-polarized. The case without strong correlations (with the spin-independent masses, SIM) is analyzed for comparison. We consider both the $s$-wave and the $d$-wave symmetries of the superconducting gap and concentrate on the parallel orientation of the Cooper pair momentum $\bQ$ with respect to the junction interface. The junction conductance is presented for selected barrier strengths (i.e., in the contact, intermediate, and tunneling limits). The conductance spectra in the cases with and without strong correlations differ essentially. Our analysis provides thus an experimentally accessible \textit{test for the presence of strong-correlations in the superconducting state}. Namely, correlations alter the distance between the conductance peaks (or related conductance features) for carriers with spin-up and spin-down. In the uncorrelated case, this distance is twice the Zeeman energy. In the correlated case, the corresponding distance is about 30-50$\%$ smaller, but other models may provide even stronger difference, depending on details of the system electronic structure. It turns out that the strong correlations manifest themselves most clearly in the case of the junction with the BCS, rather than the FFLO superconductor, what should make the experimental verification of the present results simpler.
\end{abstract}

\pacs{74.45.+c, 71.27.+a, 71.10.Ca, 74.50.+r}


\maketitle

\section{Introduction}

The search for evidence of strong electron correlations in condensed matter has concentrated in recent years on superconducting state in unconventional materials and its coexistence with magnetism. One of such examples is the search for experimental evidence for the Fulde-Ferrell-Larkin-Ovchinnikov (FFLO) superconducting state. The FFLO state was proposed theoretically in the 1960s. \cite{FF,LO,*LO2} In this unconventional superconducting state the Fermi wave vector difference for the electrons with spin-up and -down due to the presence of Zeeman term makes it favorable for the Cooper pair to acquire a nonzero total momentum $\bQ = 2 \bq$. Consequently, the phase of the superconducting gap parameter oscillates spatially with the wave vector $\bQ$. By forming such a condensate of moving Cooper pairs, the superconducting state persists to magnetic fields remarkably higher than the Pauli $H_{c2}$ limit. The FFLO state has suddenly gained renewed interest recently (for a review see Ref.~ \onlinecite{Matsuda-Rev}) because of its possible detection in the heavy fermion superconductor CeCoIn$_5$, \cite{Kakuyanagi, Kumagai2, Bianchi, Kumagai} although the nature of the high-field low-temperature phase observed in this system is still under an intensive debate after antiferromagnetism has been observed in the vicinity of this phase. \cite{Young, Kenzelmann, Koutroulakis, Ikeda, Hatakeyama} The FFLO state has also been proposed for $\kappa$-(BEDT-TTF)$_2$Cu(NCS)$_2$, \cite{Singleton, Bergk} $\beta''$-(ET)$_2$SF$_5$CH$_2$CF$_2$SO$_3$, \cite{Cho} and other layered organic superconductors (see References in Ref.~\onlinecite{Bergk}). Also, existence of the FFLO state has been indicated in other heavy-fermion systems: PuRhGa$_5$, \cite{Javorsky} Ce$_2$PdIn$_8$ \cite{Dong} (see Ref.~\onlinecite{Pfleiderer}, Sec. V.B.1 for a more detailed account), as well as in the pnictide superconductor LiFeAs. \cite{Cho2} The FFLO state has also been investigated in high density quark and nuclear matter, \cite{Casalbuoni} as well as in optical lattices.~\cite{Machida,Kinnunen,Koponen}

All the systems considered so far to be a host to the FFLO phase have a \textit{reduced dimensionality}, what is crucial for the FFLO phase stability, as then the orbital effects are suppressed and the Pauli effect (Zeeman splitting) may become the dominant factor. Another obvious feature, which suppresses the orbital effects, is the \textit{heavy quasiparticle mass}. These characteristics of possible FFLO hosts indicate that these systems are \textit{likely to have strong electron (fermion) correlations} and thus also possess specific features resulting from them.


The role of strong correlations in the most likely candidate for the FFLO state, CeCoIn$_5$ is essential not only because this system is a heavy fermion superconductor, with very narrow bands originating from $4f$ electrons hybridized with $5d-6s$ states. What is equally important, the spin-dependent effective masses (SDM) of quasiparticles have been directly observed in this system \cite{McCollam} by means of the de Haas-van Alphen oscillations in a strong applied magnetic field. SDM are one of the hallmarks of strong correlations, as they appear naturally in theories incorporating correlations (Gutzwiller, \cite{JS_Gopalan} slave-bosons, \cite{Korbel, SBREVIEW}
dynamical mean field theory, \cite{Bauer} fluctuation-exchange approximation \cite{Onari}), when the system is spin-polarized.\footnote{SDM have also been observed in other heavy-fermion systems. \cite{Sheikin, Takashita} One should note at the outset that by spin dependent masses (or their enhancement) we understand a $\bk$-independent feature of a very narrow band which is derived solely from the effect of correlation. The ordinary splitting into spin subbands in the Zeeman field is a separate effect.}

Because of the above reasons, it is important to study the effect of correlations on the FFLO phase. Such analysis has already been performed in a few cases, \cite{JKJS,JKJS2,Maska,Acta,Ying} and it indicates, among others, that the interelectronic correlations play an important role in forming and stabilizing the FFLO phase.

In the present paper we concentrate on providing an experimentally-accessible concrete characteristics of a superconducting state with strong correlations. Namely, we study conductance of a normal metal - superconductor junction (NSJ) with the strongly-correlated superconductor in either the Fulde-Ferrell (FF) type of the FFLO state, or the Bardeen-Cooper-Schrieffer (BCS) state (the latter is stable in lower fields). Conductance spectroscopy of such junction is an experiment sensitive to both the phase and amplitude modulation of the superconducting order parameter, and therefore it is a candidate technique for providing a direct evidence for the presence of the FFLO phase.
In that situation, a crucial role is played by the Andreev reflection (AR) processes. \cite{Andreev,*Andreev2} In the simplest view of the Andreev reflection, an incident electron entering from the normal metal into the superconductor (SC) is converted at the NSJ interface into a hole moving in the opposite direction (to the incident particle) and Cooper pair inside SC. Such processes increase conductance of the junction (in an ideal case by a factor of two), which is analyzed in the framework provided by Blonder, Tinkham, and Klapwijk. \cite{BTK}

The conductance characteristics for a NSJ with superconductor in the FFLO state has already been investigated for both the cases of the FF (with $\Delta(\bfr) = \Delta_\bQ e^{i\bQ \bfr}$) \cite{Cui, JKC, Partyka} and the Larkin-Ovchinnikov ($\Delta(\bfr) = \Delta_\bQ \cos(\bQ \bfr)$) \cite{Tanaka2} types of FFLO states, as well as for the case of superconductor with a supercurrent \cite{Zhang, Lukic} (i.e. the situation similar to ours from formal point of view). See also Refs.~\onlinecite{Tanaka1, Bruder, Kashiwaya, Argyropoulos} for the case of NSJ with BCS state of the $d$-wave symmetry. None of the above papers have taken into account strong electron correlations.

Here we consider both the cases of $s$-wave and $d$-wave strongly-correlated superconductor in magnetic field and in the Pauli limiting situation (i.e., we neglect orbital effects, as the Maki parameter \cite{Maki} in the systems of interest is quite high \cite{Kumagai2}). The strong correlations are taken into account by assuming dispersion relations with SDM of quasiparticles and with the correlation field, as given e.g. by the Gutzwiller approximation, \cite{JS_Gopalan} or slave-boson theory. \cite{Korbel} The case without strong correlations (with spin-independent masses, SIM) is analyzed for comparison. In low magnetic fields the superconductor is in the BCS state, and in higher magnetic fields a transition to the FFLO state takes place. We consider only the \tr{simpler} FF type of FFLO state \tr{as we intend to single-out novel features of the situation with strong correlations in the simplest case (the analysis of LO state is much more complex \cite{Tanaka2}). Our study already leads to interesting, novel results}. We set the direction of the Cooper pair momentum $\bQ$ as either perpendicular, or parallel to the junction interface, with more attention paid to the latter situation. The analysis is performed in a fully self-consistent manner. Namely, we select Cooper pair momentum $\bQ$ minimizing the free energy of the system and we determine the chemical potential $\mu$ in each phase separately so that the particle number $n$ is kept constant. Such an adjustment of $\mu$ is required even for the BCS state for the narrow-band case. Also, such a careful examination of the superconductor properties is important, and non-self-consistent calculations may lead to important alterations of the conductance spectrum. \cite{JKC}

As we deal with heavy quasiparticles on the superconducting side of NSJ, we should in principle take into account the Fermi-velocity-mismatch effects. Under those circumstances, the AR processes would be severely limited by a high effective barrier strength $Z$. On the other hand, AR is clearly observed in junctions with heavy-fermion superconductors \cite{Park, Goll} and theoretical efforts have been made to understand why this is the case. \cite{Deutscher, Araujo, Araujo2} Based on these studies, we disregard the Fermi-velocity mismatch by assuming equal chemical potentials and equal \textit{average} masses of quasiparticles on both sides of the junction. Namely, we choose masses on the normal side as $m_{av}$, and on the superconductor side we have that $(m_\ua+m_\da)/2 = m_{av}$, with $m_{av} = 100 \textrm{ m}_0$ \tr{(where $\textrm{ m}_0$ is the electron mass in vacuum)}, which roughly corresponds to the heaviest band of CeCoIn$_5$. \cite{McCollam} This assumption is, in our view, a justifiable simplification, as we would like to single out the novel features in their clearest form. Note also, that we consider a model situation with its parameters taken from the experiment for CeCoIn$_5$.

In brief, we study conductance of NSJ with superconductor exhibiting strong electron correlations (SDM case). To single out novel features of such situation, we also study the uncorrelated case (SIM) and compare those results.

The paper is organized as follows. In Sec. \ref{sec:sc} we discuss the superconducting state of quasiparticles with SDM and SIM for a two-dimensional electron gas. In Sec. \ref{sec:AR} we present the theory concerning conductance of a normal metal - strongly-correlated superconductor junction. In Sec. \ref{sec:results} we show conductance spectra for the cases with SDM and SIM. In Sec. \ref{sec:exp} we discuss relation of our results to experiments and suggest their possible experimental verification. Finally, in Sec. \ref{sec:summary} we provide a brief summary.

\section{Fulde-Ferrell superconducting state basic characteristics: model and method} \label{sec:sc}

As said above, here we consider a two-dimensional system of paired quasiparticles in the situations with SDM and SIM. The system of self-consistent equations describing such superconducting state has already been presented in detail in Refs.~\onlinecite{JKJS, JKJS2}. For the sake of completeness, we provide here a brief summary of our procedure. We start with the Hamiltonian
\eq
\hat{\mathcal{H}} = \sum_{\bk \sigma} \xi_{\bk \sigma} a^\dagger_{\bk \sigma} a_{\bk \sigma} + \frac{1}{N} \sum_{\bk \bk' \bq} V_{\bk, \bk'} a^\dagger_{\bk+\bq \uparrow} a^\dagger_{-\bk + \bq \downarrow} a_{-\bk' + \bq \downarrow} a_{\bk' + \bq \uparrow} + \frac{N}{n} \overline{m} h_{cor}, \label{eq:HStart}
\eqx
where $\bQ=2\bq$ is the wave vector of the Cooper pair center of mass, \tr{$n \equiv n_\ua +n_\da$ is the band filling, $\overline{m} \equiv n_\ua - n_\da$ is the spin-polarization of the system, and $N$ is the total number of particles.} The dispersion relation for the cases with SDM and SIM is chosen, respectively, as
\eqn
\xi_{\bk \sigma} &=& \frac{\hbar^2 k^2}{2 m_\sigma} - \sigma (h + h_{cor}) - \mu, \label{eq:dispSDM}\\
\xi_{\bk \sigma}^{(SIM)} &=& \frac{\hbar^2 k^2}{2 m_{av}} - \sigma h - \mu, \label{eq:dispSIM}
\eqnx
where $h\equiv g \mu_B H/2$ with $H$ being the applied magnetic field. The quantity $h_{cor}$ is the correlation field which appears naturally in both the slave-boson theory (it is equivalent to $- \beta$ of Ref.~\onlinecite{Korbel}) and Gutzwiller approximation if this approximation is performed with care. \cite{SGA,Wang,Yang} Justification of a Hamiltonian with both the pairing part and SDM can be found in Ref.~ \onlinecite{Maska} (Appendix A) and in Ref.~\onlinecite{JS_RSP1,*JS_RSP2,*JS_RSP3}. The spin-dependent quasiparticle mass is equal to $m_\s \equiv m_B / q_\s(n, \overline{m})$, where $m_B$ is the bare band mass and $q_\s(n, \overline{m})$ is the band-narrowing factor. Explicitly (in the Hubbard $U \to \infty$ limit) the quasiparticle masses are given by \cite{JS_Gopalan, Korbel}
\begin{equation}
\frac{m_\sigma}{m_B} = \frac{1-n_\sigma}{1-n} = \frac{1-n/2}{1-n} - \sigma \, \frac{\overline{m}}{2(1-n)} \equiv \frac{1}{m_B} (m_{av} - \sigma \Delta m/2), \label{eq:m}
\end{equation}
with $\Delta m \equiv m_\da - m_\ua$. Next, as in the BCS theory, we take the pairing potential in a separable form and assume it is nonzero in a small region around the Fermi surface (for details see Refs.~\onlinecite{JKJS}, \onlinecite{JKJS2}, and \onlinecite{JKPHD})
\eq
V_{\bk, \bk'} = - V_0 \eta_\bk \eta_{\bk'},
\eqx
where $\eta_\bk \equiv \cos{(a k_x)} - \cos{(a k_y)}$ for the $d$-wave case (with $a = 4.62\textrm{ \AA}$ being the lattice constant for CeCoIn$_5$ \cite{Petrovic}) and $\eta_\bk \equiv 1$ for the $s$-wave case. Under such assumptions, the superconducting gap can be factorized as
\eq
\Delta_{\bk, \bQ} = \Delta_\bQ \eta_\bk. \label{eq:delta}
\eqx
Following the standard mean-field approach to Hamiltonian (\ref{eq:HStart}), we obtain the generalized free-energy functional $\mathcal{F}$ and the system of self-consistent equations as follows \cite{JKJS, JKJS2}
%
\eqn
 \mathcal{F} &=& - k_B T \sum_{\bk \sigma} \ln (1 + e^{-\beta E_{\bk \sigma}}) + \sum_\bk (\xi^{(s)}_\bk - E_\bk) + N \frac{\Delta_\bQ^2}{V_0} + \mu N + \frac{N}{n} \overline{m} h_{cor}, \label{eq:sc1} \\
 h_{cor} &=& - \frac{n}{N} \sum_{\bk \sigma} f(E_{\bk \sigma}) \frac{\partial E_{\bk \sigma}}{\partial \overline{m}} + \frac{n}{N} \sum_\bk \frac{\partial \xi_\bk^{(s)}}{\partial \overline{m}} \Big( 1 - \frac{\xi_\bk^{(s)}}{E_\bk} \Big), \label{eq:sc2} \\
 \overline{m} &=& \frac{n}{N} \sum_{\bk \sigma} \sigma f(E_{\bk \sigma}), \label{eq:sc3}\\
 \Delta_\bQ &=& \frac{V_0}{N} \sum_\bk \eta_\bk^2 \frac{1 - f(E_{\bk \uparrow}) - f(E_{\bk \downarrow})}{2 E_\bk} \Delta_\bQ, \label{eq:sc4} \\
 n &=& n_\uparrow + n_\downarrow = \frac{n}{N} \sum_{\bk \sigma} \{u_\bk^2 f(E_{\bk \sigma}) + v_\bk^2 [1 - f(E_{\bk, -\sigma})]\}, \label{eq:sc5}
\eqnx
where $\mathcal{F}(T, H, \mu; \overline{m}, h_{cor}, \Delta_\bQ, n)$ is the system free-energy functional for the case of a fixed number of particles \cite{Koponen} (we fix the band filling at the value $n=0.97$), $V_0$ is the interaction potential, $u_\bk$, $v_\bk$ are the Bogolyubov coherence coefficients, $f(E_{\bk \sigma})$ is the Fermi distribution, and $n_\sigma$ is the spin-subband filling. The physical solution is that with a particular $\bQ$ which minimizes the free energy $F$, which in turn is obtained from $\mathcal{F}$ by evaluating the latter at the values of parameters being solution to Eqs.~(\ref{eq:sc2})-(\ref{eq:sc5}). The state with $\bQ = 0$ is called the BCS state, and that with $\bQ \neq 0$ - the FF state.

The quasiparticle spectrum in the paired state is characterized by the energies (cf. also Ref.~\onlinecite{Shimahara})
\eqn
E_{\bk \sigma} & \equiv & E_\bk + \sigma \xi^{(a)}_\bk, \quad \quad \quad \quad \quad E_\bk \equiv \sqrt{\xi^{(s)2}_\bk + |\Delta_{\bk, \bQ}|^2}, \label{eq:Ek1} \\
 \xi^{(s)}_\bk & \equiv & \frac{1}{2} (\xi_{\bk + \bq \uparrow} + \xi_{-\bk + \bq \downarrow}), \quad  \xi^{(a)}_\bk \equiv \frac{1}{2} (\xi_{\bk + \bq \uparrow} - \xi_{-\bk + \bq \downarrow}). \label{eq:Ek}
\eqnx
Eqs.~(\ref{eq:sc2})-(\ref{eq:sc5}) are solved by numerical integration over the reciprocal space. We use procedures from GNU Scientific Library \cite{GSL} as solvers. For the SIM case $h_{cor}=0$ and we solve only Eqs.~(\ref{eq:sc3})-(\ref{eq:sc5}). The numerical procedure has been elaborated in detail elsewhere. \cite{JKPHD} \trr{Here, for completeness, we also provide in Tables I and II the numerical values of selected parameters for the situations with the $s$-wave and the $d$-wave symmetries of the superconducting gap, respectively}. The quantity $F_{NS}$ is the free energy of the normal state, and therefore $\Delta F$ is the condensation energy. Also, $\Delta m \equiv m_2-m_1$ is the mass difference and $h_{cor\,FS}$ is the correlation field value in the normal state. The free energies are calculated per elementary cell. The numerical accuracy is not smaller than on the level of the last digit specified.

\begin{center}
\begin{tabular}{| c | c || c | c |}
\hline
  \multicolumn{4}{|c|}{Table I. Equilibrium values of mean-field variables and related quantities} \\
  \multicolumn{4}{|c|}{for the $s$-wave solution with $H = 10.01 \textrm{ T}$ and $T=0.02 \textrm{ K}$.} \\
\hline
Variable & Value & Variable & Value  \\ \hline
$ \overline{m} $                  & 0.0129431 	       & $ \Delta m \,(m_0)$                            & 2.51322 \\
$ h_{cor} \,(\mathrm{K}) $        & -3.08230           & $ h_{cor\,FS}\,(\mathrm{K}) $ 		            & -3.26546 \\
$ \Delta_{\bQ} \,(\mathrm{K}) $   & 1.38922 	       & $ |\bQ|\,(\textrm{\AA}^{-1}) $                 & 0.00947 \\
$ \mu \,(\mathrm{K}) $            & 126.287            & $ |\bQ| / \Delta k_F $                         & 1.08 \\
$ F \,(\mathrm{K}) $              & 61.18200288        & $ \Delta F \,(\mathrm{K}) \equiv F_{NS} - F $  & -0.00111351  \\ \hline
\end{tabular}
\end{center}

\begin{center}
\begin{tabular}{| c | c || c | c |}
\hline
  \multicolumn{4}{|c|}{Table II. Equilibrium values of mean-field variables and related quantities} \\
  \multicolumn{4}{|c|}{for the $d$-wave solution with $H = 20.01 \textrm{ T}$ and $T=0.1 \textrm{ K}$.} \\
\hline
Variable & Value & Variable & Value  \\ \hline
$ \overline{m} $                  & 0.0268690 	       & $ \Delta m \,(m_0)$                            & 5.21729 \\
$ h_{cor} \,(\mathrm{K}) $        & -6.40870           & $ h_{cor\,FS}\,(\mathrm{K}) $ 		            & -6.53133 \\
$ \Delta_{\bQ} \,(\mathrm{K}) $   & 1.27455 	       & $ |\bQ|\,(\textrm{\AA}^{-1}) $                 & 0.0183 \\
$ \mu \,(\mathrm{K}) $            & 126.416            & $ |\bQ| / \Delta k_F $                         & 1.15 \\
$ F \,(\mathrm{K}) $              & 61.04342125        & $ \Delta F \,(\mathrm{K}) \equiv F_{NS} - F $  & -0.00142436 \\ \hline
\end{tabular}
\end{center}

\tr{The input parameters in our method have the following values: the band filling $n=0.97$, the lattice constant $a = 4.62\textrm{ \AA}$, the interaction potential strength $V_0/n=90 \textrm{ K}$ ($d$-wave) and $V_0/n=110 \textrm{ K}$ ($s$-wave), the interaction potential width (cutoff) $\hbar \omega_C = 17 \textrm{ K}$, the quasiparticle average mass $m_{av} = 100 \textrm{ m}_0$. The other parameters (in particular: $\overline{m}$, $h_{cor}$, $\Delta_\bQ$, $\mu$, $\bQ$, and $\theta_\bQ$) are determined from the solution procedure.}

\begin{figure}
  \begin{center}
  {\includegraphics[width=12cm,angle=270]{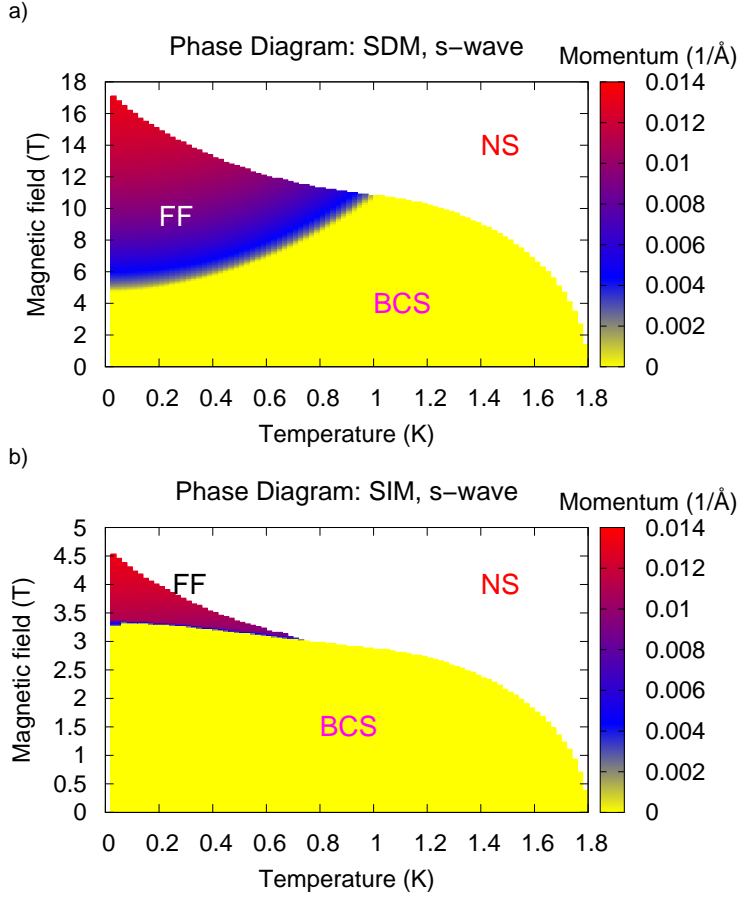}}
  \end{center}
  \caption{(Color online) Phase diagram for the $s$-wave gap symmetry in the SDM (a) and SIM (b) cases. Light (yellow) region corresponds to $\bQ = 0$ (BCS phase), the darker to the state with $\bQ \neq 0$ (FF phase) and the white to normal state (NS). Note the greater difference between SDM and SIM cases than for $d$-wave gap symmetry (see Fig.~\ref{fig2}).}
  \label{fig1}
\end{figure}

\begin{figure}
  \begin{center}
 \hspace{0cm}\includegraphics[width=13cm,angle=270]{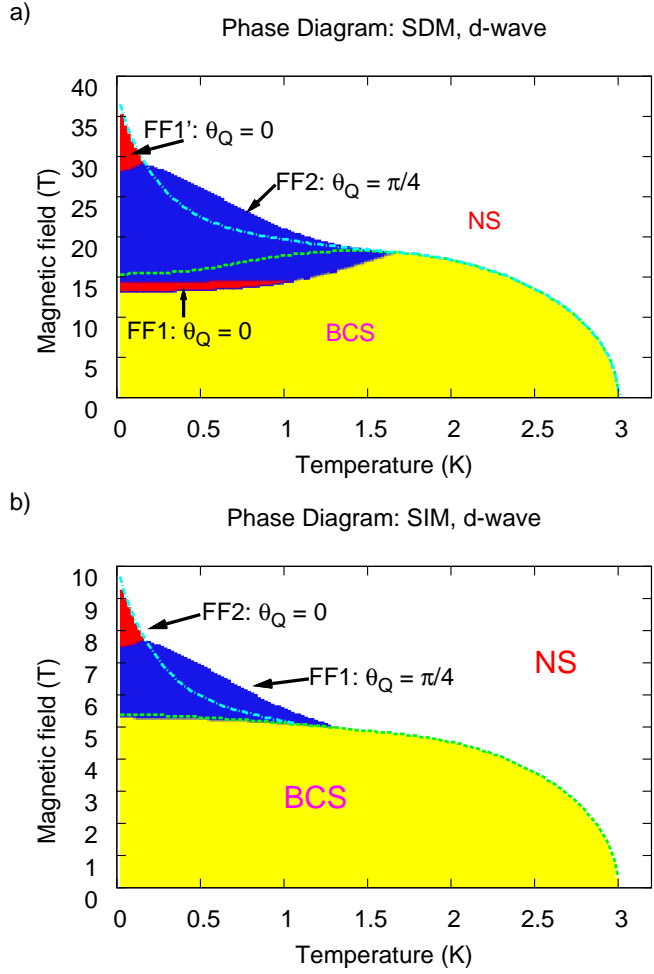}
  \end{center}
  \caption{(Color online) Phase diagram for the $d$-wave gap symmetry for the cases with SDM, (a) and SIM, (b). Light (yellow) region corresponds to $\bQ = 0$ (BCS phase), the darker (blue, red) to the states with $\bQ \neq 0$ (FF phases) and the white to normal state (NS). The red regions correspond to the Cooper-pair momentum $\bQ$ in the \trr{maximum-gap (antinodal)} direction ($\theta_\bQ = 0$), whereas the blue ones to the momentum along the \trr{nodal} direction ($\theta_\bQ = \pi/4$). Note that this anisotropy results solely from the $d$-wave gap symmetry, as the unpaired gas is isotropic. The dashed line marks the BCS critical field $H_{c2}$ in the Pauli limit, and the dot-dashed line marks $H_{c2}$ for the solution with $\theta_\bQ = 0$.}
 \label{fig2}
\end{figure}

Exemplary phase diagrams obtained on the applied field $H$ and temperature $T$ plane are exhibited in Figs.~\ref{fig1} and~\ref{fig2} for the $s$-wave and the $d$-wave cases, respectively. \trr{The angle $\theta_\bQ$ is the angle between the maximum-gap (antinodal) direction and the Cooper pair momentum $\bQ$.} \tr{Note that in both situations the FF state is more robust (i.e. the FF state fills a wider field-temperature range on the phase diagram) in the SDM case than in the SIM case. The mechanism of the FF state stabilization by strong correlations has been analyzed in detail in Refs. \onlinecite{JKJS, JKJS2, Maska}. For the sake of completeness, let us mention that this mechanism is based on a smaller Fermi wave vector splitting ($\Delta k_F \equiv k_{F\ua} - k_{F\da}$) in the SDM situation. In such case, the system can resist more efficiently the destabilizing influence of the applied magnetic field (hence higher critical fields in the SDM case). Also, it turns out that the FF state can benefit to a greater extent than the BCS state from the smaller $\Delta k_F$, as the FF state has higher spin-polarization, which is necessary for the appearance of SDM (for details see Refs. \onlinecite{JKJS}, \onlinecite{JKJS2}, and \onlinecite{JKPHD}).}

For further analysis of the Andreev reflection, we take the parameters obtained along the $T=0.02 K \approx 0$ line in Figs. \ref{fig1} and \ref{fig2}. \tr{Therefore, the results will have, strictly speaking, practical relevance for $T \ll T_{sc}$, with the superconducting transition temperature $T_{sc}~\approx~2~-~3~\textrm{ K}$, as can be seen from Figs. \ref{fig1} and \ref{fig2}.}

\section{Junction conductance: theoretical analysis} \label{sec:AR}

\begin{figure}
\includegraphics[width=14cm]{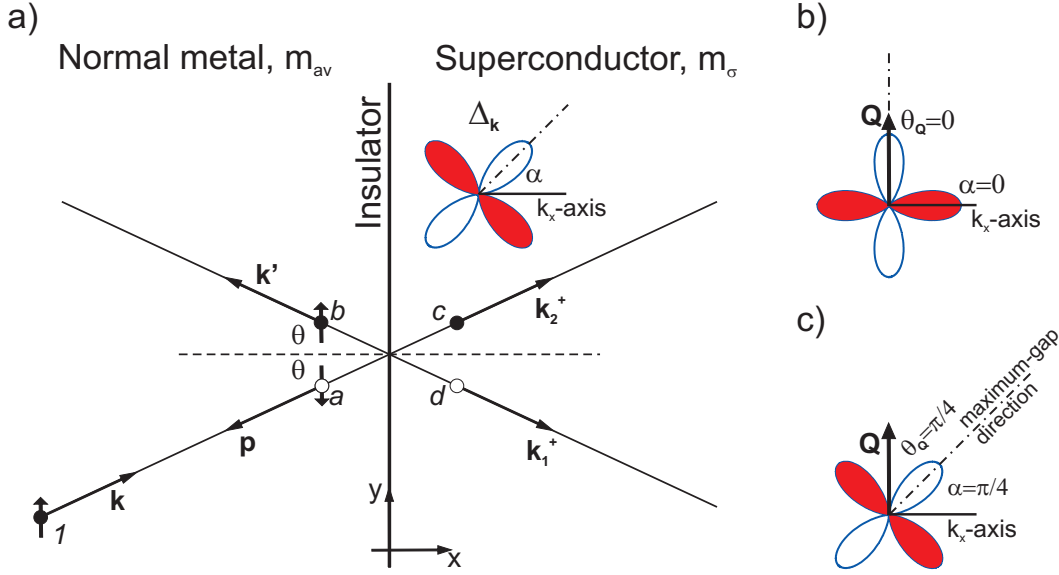}
\caption{\label{fig3} (Color online) a) Junction geometry for incoming particle of spin $\sigma = \ua$. Normal-state and superconductor regions are marked. Interface lies at the $x=0$ line. The superconducting gap is also presented: $\alpha$ is the angle between the $k_x$ axis and maximum-gap direction. Full circles mark quasiparticles and empty ones mark quasiholes. Momentum of each of them is marked with a boldface letter, and amplitude with an italic letter. Namely, incoming particle has the momentum $\bk$, and amplitude 1, reflected hole has $\bf{p}$ and ${\it a}$, reflected quasiparticle: $\bk'$, ${\it b}$, transmitted quasiparticle: $\bf{k_2^+}$, ${\it c}$, and transmitted quasihole: $\bf{k_1^+}$, ${\it d}$. The angle of incidence is equal to $\theta$ and to the angle of reflection but other angles (of reflection of quasihole and those of transmissions) may differ (cf. also Fig.~\ref{fig4}). \trr{In (b) and (c) we show explicitly the two $d$-wave configurations of the superconducting gap for the FF phase studied in the following: (b) corresponds to Fig.~\ref{fig7} and (c) to Fig.~\ref{fig8}.}}
\end{figure}

For the analysis of the NSJ conductance we take the superconducting state parameters obtained self-consistently (from the procedure presented above). We consider here only two-dimensional NSJ for simplicity. Kinematics of the reflection may be analyzed by means of the Bogolyubov-de Gennes (BdG) equations \cite{BdG}
\eqn
E u_\s(\mb{x}) &=& \hat{\mathcal{H}}_0 u_\s(\mb{x}) + \int d\mb{x}' \Delta(\mb{s}, \mb{r}) v_\sigma (\mb{x}'), \label{eq:BdG1} \\
E v_\s(\mb{x}) &=& -\hat{\mathcal{H}}_0 v_\s(\mb{x}) + \int d\mb{x}' \Delta^*(\mb{s}, \mb{r}) u_\sigma (\mb{x}'), \label{eq:BdG2}
\eqnx
where $\mb{s} = \mb{x} - \mb{x}'$, $\bfr = (\mb{x}+\mb{x}')/2$, and $\s = \pm 1$ is the spin quantum number of the incoming quasiparticle and $u_\s(\mb{x})$ and $v_\s(\mb{x})$ are the particle and hole wave-function components. The one-particle Hamiltonian is given by
\eq
\hat{\mathcal{H}}_{0}(\bfr) = - \nabla \frac{\hbar^2 }{2 \, m(\bfr)} \nabla - \sigma h - \s h_{cor}(\bfr)- \mu + V(\bfr),
\eqx
where we have used the effective mass approximation \cite{Burt, Mortensen} to express the kinetic part as $\nabla \frac{\hbar^2 }{2 \, m(x)} \nabla$ with $m(\bfr) \equiv m(x) = m_{av} \Theta(-x) + m_\sigma \Theta(x)$, similarly as in Refs.~ \onlinecite{Annunziata,Annunziata2,Annunziata3,Mortensen}. The correlation field is nonzero only on the superconducting side of the junction ($h_{cor}(\bfr) = h_{cor} \Theta(x)$). Also, $\bfr = (x, y)$ and the interface scattering potential is chosen as a delta function of strength $\tilde{H}$, i.e. $V(\bfr)~=~\tilde{H}~\delta(x)$. The gap function can be Fourier transformed as following
\eq
\Delta(\mb{s}, \mb{r}) = \int d\bk e^{i \bk \mb{s}} \tilde{\Delta}(\bk, \bfr) = \int d\bk e^{i \bk \mb{s}} \Delta_{\bk, \bQ} \, e^{i \bQ \bfr}\, \Theta(x), \label{eq:DeltaFT}
\eqx
with $\Delta_{\bk, \bQ}$ \trr{as in Eq.~(\ref{eq:delta}) but with the original set of coordinates rotated by $\alpha$ (cf. Fig.~\ref{fig3}). Explicitly, the superconducting gap we use from now on has the form (in the new coordinates)
\eq
\Delta_{\bk, \bQ} = \Delta_\bQ \Big(\cos{(a k_x \cos{\alpha} - a k_y \sin{\alpha} )} - \cos{(a k_y \cos{\alpha} + a k_x \sin{\alpha})}\Big).
\eqx}
We neglect the proximity effects by assuming a step-like gap function. To solve the BdG equations we make the plane-wave ansatz. Namely, we assume that the two-component pair wave function has the form
\eq
\psi(\bfr, \sigma) \equiv \left( \begin{array}{c} u_\s(\bfr) | \sigma \rangle \\ v_\s(\bfr) |\overline{\sigma}\rangle \end{array} \right) = e^{i \bk \bfr} \left( \begin{array}{c} \tilde{u} e^{i \bq \bfr} \, | \sigma \rangle \\ \tilde{v} e^{-i \bq \bfr} \, |\overline{\sigma}\rangle \end{array} \right), \label{eq:psi1}
\eqx
with $\tilde{u}$ and $\tilde{v}$ as constants and with $\overline{\sigma} \equiv -\sigma$ (we have also dropped the $\s$ indices of $\tilde{u}$ and  $\tilde{v}$). We also remind that $\bq = \bQ/2$. By substituting (\ref{eq:DeltaFT}) and (\ref{eq:psi1}) into BdG equations (\ref{eq:BdG1}), (\ref{eq:BdG2}) and after some algebra we obtain the following matrix equation
\eq
\left( \begin{array}{cc}
-E + \xi_{\bk+\bq, \sigma} & \Delta_{\bk, \bQ} \\
\Delta_{-\bk, \bQ}^* & -E - \xi_{\bk-\bq, \overline{\sigma}}
\end{array} \right) \left( \begin{array}{c} \tilde{u} \, | \sigma \rangle \\ \tilde{v} \, |\overline{\sigma}\rangle \end{array} \right) =  0, \label{eq:matrix}
\eqx
where unpaired quasiparticle energies $\xi_{\bk \s}$ are given by Eq. (\ref{eq:dispSDM}) or (\ref{eq:dispSIM}). Eq.~(\ref{eq:matrix}) gives the dispersion relations for quasiparticles and quasiholes in the superconductor
\eq
E = E_{\bk\pm} = \left\{ \begin{array}{c} \xi_\bk^{(a)} \pm \sqrt{\xi_\bk^{(s)2} + \Delta_{\bk, \bQ} \Delta^*_{-\bk, \bQ}} \quad \, \textrm{for } \sigma = \ua, \\
- \xi_{-\bk}^{(a)} \pm \sqrt{\xi_{-\bk}^{(s)2} + \Delta_{\bk, \bQ} \Delta^*_{-\bk, \bQ}} \quad \textrm{for } \sigma = \da, \end{array} \right.
\label{eq:Epm}
\eqx
where $\xi_\bk^{(s,a)}$ have been defined in Eq.~(\ref{eq:Ek}). One may check that the above equation is in accordance with Eq.~(\ref{eq:Ek1}), as $E_{\bk+} = E_{\bk\ua}$ (quasiparticle) and $E_{\bk-} = - E_{\bk\da}$ (quasihole) for incoming particle with spin $\sigma = \ua$, as well as $E_{\bk+} = E_{-\bk\da}$ (quasiparticle) and $E_{\bk-} = - E_{-\bk\ua}$ (quasihole) for incoming particle with spin $\sigma = \da$. This holds as long as $\Delta^*_{-\bk, \bQ} = \Delta^*_{\bk, \bQ}$, which is true for any real $\bk$.

As already mentioned, we study the FF type of the FFLO superconducting state, in which $\Delta(\bfr)~=~\Delta_\bQ e^{i 2 \bq \bfr}$ and set the direction of the Cooper pair momentum $\bQ = 2\bq$ as either perpendicular ($\bQ = (Q, 0)$), or parallel ($\bQ = (0, Q)$) to the junction interface. The \tr{perpendicular} configuration ($\bQ = (Q, 0)$) may lead to \tr{accumulating of charge at the NSJ interface due to normal and/or supercurrent present in the FF state}. Therefore we pay principal attention to the \tr{parallel} configuration. Parenthetically, the accumulation processes are very slow for the case of heavy quasiparticles.

\begin{figure*}
\includegraphics[width=15cm]{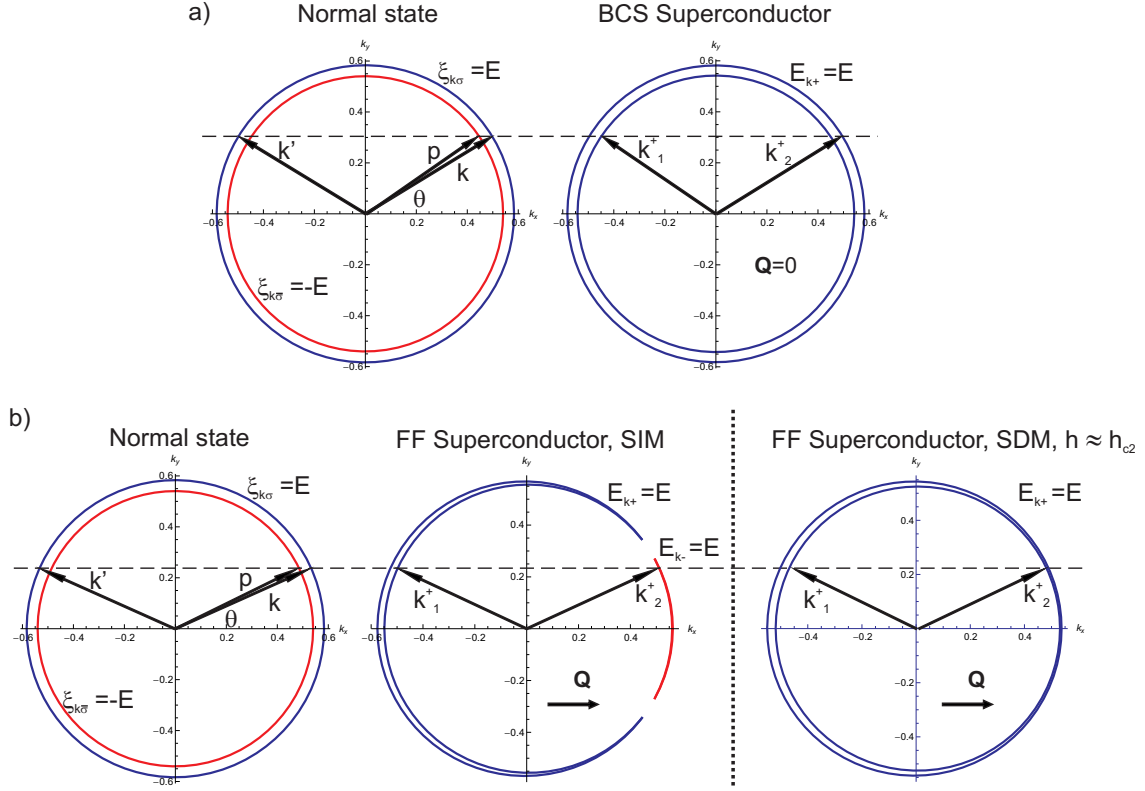}
\caption{\label{fig4} (Colour online). The junction geometry in reciprocal space. All relevant vectors are marked. It can be seen that only the incident and reflection angles are equal to $\theta$. It can be anticipated at this point that changing $\theta$ for BCS state does not lead to drastic changes in the transmission/reflection probabilities, whereas for the FF state the situation is quite different since $\bQ \neq 0$ induces anisotropy in the reciprocal space. The energy $E$ value has been chosen as $10 \textrm{ K}$ for all graphs except (b) "FF Superconductor" for which $E=0.01\textrm{ K} \approx 0$ (for $E>0.5 \textrm{ K}$ there would be no $E=E_{\bk-}$ regions in this case). The dashed lines are guide to eye and illustrate the conservation of momentum $y$-component.}
\end{figure*}

As we consider electron injected from the conductor side of the junction (junction geometry is presented in Fig.~\ref{fig3}), the corresponding wave functions can be expressed as (we have omitted the spin part for clarity)
\eqn
\psi_{<}(\bfr) = \left( \begin{array}{c} 1 \\ 0 \end{array} \right) e^{i\bk\bfr} + a \left( \begin{array}{c} 0 \\ 1 \end{array} \right) e^{i \bf{p} \bfr} + b \left( \begin{array}{c} 1 \\ 0 \end{array} \right) e^{i\bk'\bfr}, \\
\psi_{>}(\bfr) = d \left( \begin{array}{c} u_1 e^{i q_x x} \\ v_1 e^{-i q_x x} \end{array} \right) e^{ i\bk_1^{+} \bfr} + c \left( \begin{array}{c} u_2 e^{i q_x x} \\ v_2 e^{-i q_x x} \end{array} \right) e^{ i\bk_2^{+} \bfr},
\eqnx
where $\psi_{<}(\bfr)$ and $\psi_{>}(\bfr)$ describe wave function on the normal-metal and superconductor sides, respectively. The quasimomenta $\bk_1^+$ (for quasihole) and $\bk_2^+$ (for quasiparticle) are solutions of Eq.~(\ref{eq:Epm}) for a given incident energy $E$ propagating in the positive x direction. From the translational symmetry of the junction along the $y$ direction comes conservation of the $y$ momentum component. Namely, $k_y = k'_y = p_y = k_{1y}^+ = k_{2y}^+$. All the wave vectors are presented in Fig.~\ref{fig4}.

We use boundary conditions with the appropriate masses \footnote{Boundary conditions with different masses have been used before e.g. in Refs.~ \onlinecite{Annunziata,Annunziata2,Annunziata3}.} and the interface potential jump $\tilde{H}$; they are as follows
\eqn
\psi_<(\bfr)|_{x=0} = \psi_>(\bfr)|_{x=0}, \\
\frac{1}{m_{av}} \frac{\partial \psi_<(\bfr)}{\partial x}|_{x=0} = \frac{1}{m_{\sigma}} \frac{\partial \psi_>(\bfr)}{\partial x}|_{x=0} - \frac{2 \tilde{H}}{\hbar^2} \psi_<(\bfr)|_{x=0}.
\eqnx
Those conditions lead to the following set of 4 equations \footnote{These equations are written for $y=0$. If $y\neq 0$ additional terms $e^{\pm i q_y y}$ appear, but they do not alter the solution, so they are usually omitted for clarity.} for the amplitudes ($a, b, c, d$)
\begin{widetext}
\eqn
1 + b - c u_2 - d u_1 = 0, \label{eq:set1} \\
a - c v_2 - d v_1 = 0, \\
\frac{i k_x (1 - b)}{m_{av}} - \frac{c u_2 i (q_x + k_{2x}^{+})}{m_\sigma} - \frac{d u_1 i (q_x + k_{1x}^{+})}{m_\sigma} + \frac{2 \tilde{H}}{\hbar^2}(1 + b) = 0,\\
\frac{a i p_x}{m_{av}} - \frac{c v_2 i (k_{2x}^{+} - q_x)}{m_{\overline{\sigma}}} - \frac{d v_1 i (k_{1x}^{+} - q_x)}{m_{\overline{\sigma}}} + \frac{2 \tilde{H}}{\hbar^2} a = 0, \label{eq:set4}
\eqnx
\end{widetext}
which are similar to those in e.g. Ref.~\onlinecite{Mortensen}, except in our case vectors are replaced by their $x$-components: e.g. $k \leftrightarrow k_x$, $p \leftrightarrow p_x$, and also SDM are properly accounted for (obviously in the SIM case we have that $m_\ua = m_\da = m_{av}$). From the solution of Eqs.~(\ref{eq:set1})-(\ref{eq:set4}) one can obtain probabilities of the hole reflection $p^\sigma_{rh} = |a|^2 \frac{\Re[p_x]}{k_x}$, particle reflection $p^\sigma_{re} = |b|^2$, quasiparticle transmission
\eq
p^\sigma_{te} = |c|^2 m_{av} \frac{(\frac{|u_2|^2}{m_\s} - \frac{|v_2|^2}{m_{\overline{\s}}}) \Re[k^+_{2x}] + (\frac{|u_2|^2}{m_\s} + \frac{|v_2|^2}{m_{\overline{\s}}}) q_x}{k_x},
\eqx
and quasihole transmission
\eq
p^\sigma_{th} = |d|^2 m_{av} \frac{(\frac{|u_1|^2}{m_\s} - \frac{|v_1|^2}{m_{\overline{\s}}}) \Re[k^+_{1x}] + (\frac{|u_1|^2}{m_\s} + \frac{|v_1|^2}{m_{\overline{\s}}}) q_x}{k_x},
\eqx
where the $\sigma$ superscript indicates the spin of the incoming electron. In the following we use the dimensionless barrier strength $Z~\equiv~2 m_{av} \tilde{H}/(k_F \hbar^2)$, where we define Fermi wave vector $k_F$ using the zero-field value $k_F = \frac{1}{\hbar} \sqrt{2 m_{av} \mu}$. Note also that we do not use the assumption $k = k' = p = k_1^+ = k_2^+ \approx k_F$ utilized at this point in majority of the papers on Andreev reflection spectroscopy, because we deal with heavy quasiparticles for which $\mu$ is of the order of $100 \textrm{ K}$. Therefore the usual assumption $\mu \gg E$ is not, strictly speaking, applicable in the present situation.

\section{Results and physical discussion} \label{sec:results}

Differential conductance ($G \equiv dI/dV$) can be obtained from the reflection and transmission probabilities \cite{BTK, Chaudhuri} in a straightforward manner
\eq
G_{ns}^\sigma = \frac{1}{2} \int_{-\pi/2}^{\pi/2} d\theta \cos{\theta} [1-p^\sigma_{re}(E, \theta) + p^\sigma_{rh}(E, \theta)]. \label{eq:G}
\eqx
The final result of our calculation is the total conductance $G$ averaged over spin and normalized with respect to the conductance $G_{nn}^\sigma$ of the junction with $\Delta=0$ but still with the same other parameters ($m_\s$, $\mu$, $h_{cor}$), as the superconducting state. Namely,
\eq
G = \frac{G_{ns}^\ua + G_{ns}^\da}{G_{nn}^\ua+G_{nn}^\da}.
\eqx
This quantity is exhibited in the following figures, sometimes with the spin-resolved conductance $G^\sigma~\equiv~G_{ns}^{\sigma} / G_{nn}^{\sigma}$. We assume the barrier strength equal to $Z=0$ (contact limit), $Z=0.5$ (intermediate limit), and $Z=5$ (tunneling limit).
\tr{The case of $Z=5$ reflects not only the situation for planar NSJ with a thick insulating layer, but also that encountered in Scanning Tunneling Spectroscopy (STS) experiments. \cite{Giaever,*Eschrig,*Fischer}}

Our goal in the following is to identify novel, {\it model-independent} features of the strongly-correlated situation (i.e., with SDM). Namely, those features should not depend on the assumed dispersion relation or the pairing-potential strength.

\subsection{$s$-wave pairing symmetry}

\begin{figure}[h!]
\includegraphics[width=10cm]{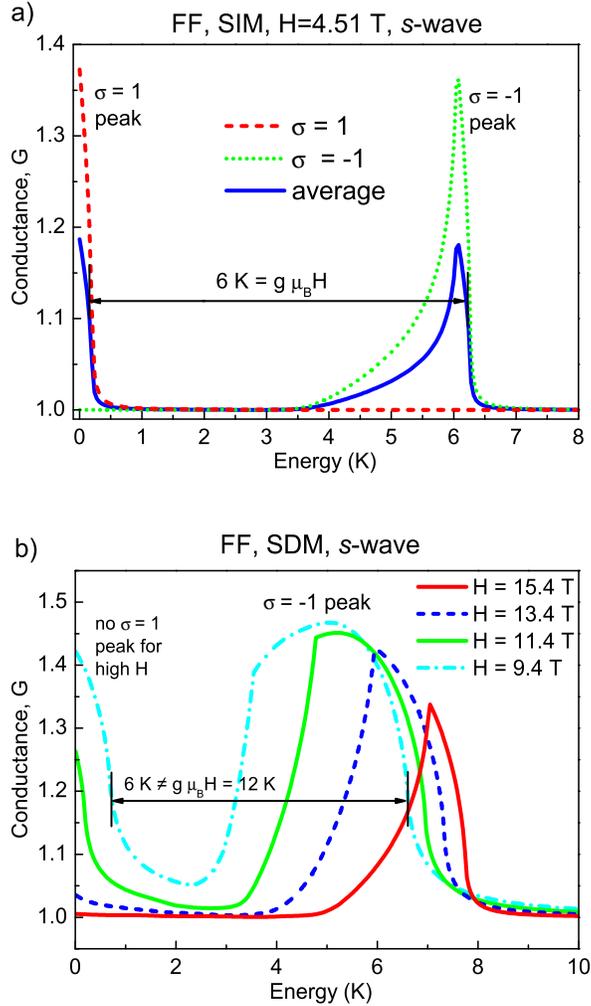}
\caption{\label{fig5} (Colour online).
Conductance spectra for the case of $s$-wave FF state for SIM (a) and SDM (b) cases. The $\bQ$ vector is oriented perpendicular to the junction and the intermediate barrier strength $Z = 0.5$ is taken. \tr{The value of the gap is (a) $\Delta_\bQ = 0.11 \textrm{ K}$; (b) $\Delta_\bQ = 0.36$, $0.70$, $1.09$, and $1.53 \textrm{ K}$ for the decreasing magnetic field.} The distance between the peaks is twice the Zeeman energy $2h = g \mu_B H$ only for the SIM case. In the SDM case the correlations compensate the Zeeman splitting (by means of $h_{cor}$ and $m_\s$), and the peaks are closer than $g \mu_B H$.
}
\end{figure}

In Fig.~\ref{fig5} the conductance for the $s$-wave gap symmetry and $\bQ$ vector oriented perpendicular to the junction, is presented. \tr{It can be seen that there are peaks in the conductance originating from AR processes of quasiparticles having different spins, that take place when the energy $E$ of the incoming electron fits into the so-called Andreev Window (AW), see Refs. \onlinecite{JKC}  (Fig. 3), \onlinecite{Partyka}, and Ref. \onlinecite{JKPHD} (Chapter 5, Figs. 5.1d, 5.4b) for more details.} These peaks are separated by a distance equal to twice the Zeeman energy ($2 h = g \mu_B H$) only in the case without strong correlations (SIM). For the SDM case the correlations compensate the Zeeman splitting (by means of $h_{cor}$ and $m_\s$, cf. Refs.~\onlinecite{JKJS, JKJS2}) and as result the conductance peaks are closer than twice the Zeeman energy. We identify this feature as a \textit{hallmark of strong correlations in the superconducting state}. Another interesting feature differentiating the SIM and SDM cases is absence of the $\s = \ua$ peak for SDM when magnetic field $H \gtrsim 12 \textrm{ T}$. For such fields the junction is transparent to incoming particles with $\s = \ua$ because the Andreev window \cite{JKC,Partyka} falls below $E=0$. In other words, the quasiparticle energy $E_{\bk+}$ within FF superconductor is below zero around the whole Fermi surface. This leads to breaking of Cooper pairs and produces normal state region filling whole angular space around the Fermi surface (see Fig.~\ref{fig4}b, SDM case). Since there are normal particles with $\s = \ua$ within the FF superconductor, the incoming $\s = \ua$ quasiparticle does not feel the superconducting gap presence, and the junction is transparent, what yields $G_\ua \approx 1$.

\begin{figure}
\includegraphics[width=13cm]{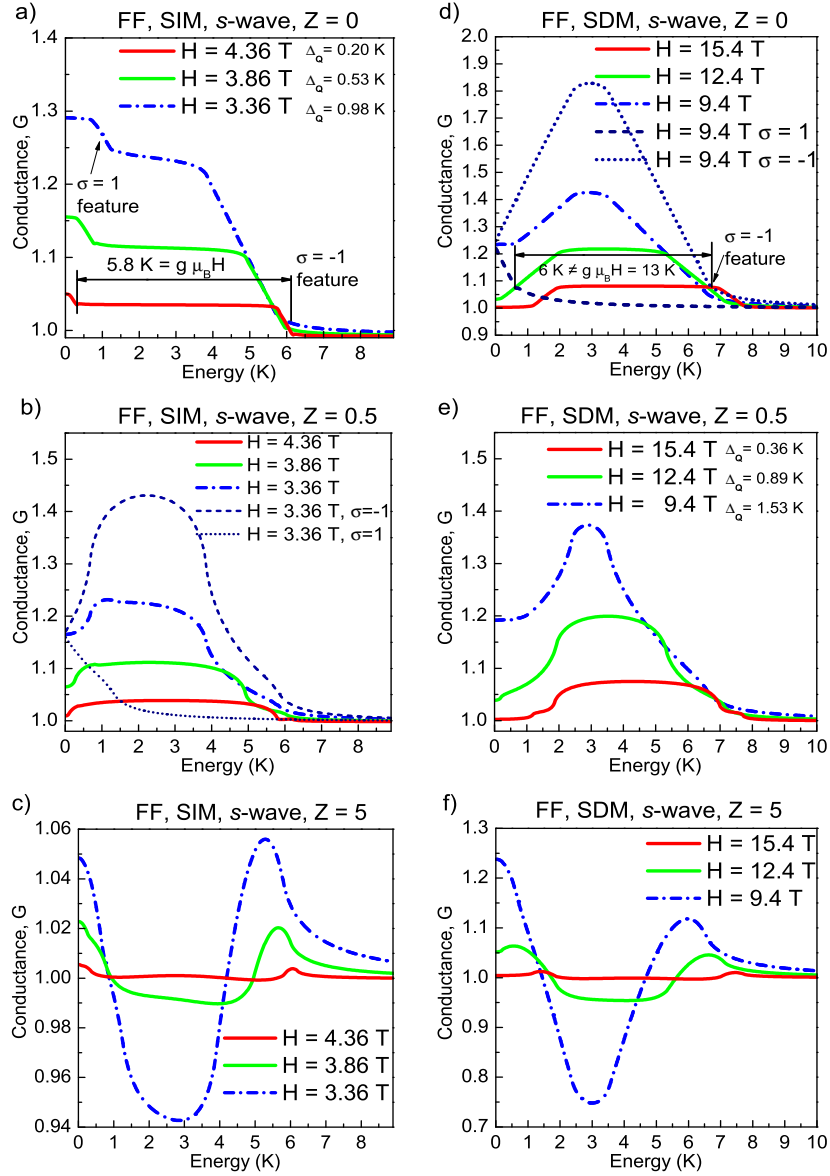}
\caption{\label{fig6} (Colour online).
Conductance spectra for the case of $s$-wave FF state for SIM (a-c) and SDM (d-f), the $\bQ$ vector oriented parallel to the junction, and selected $Z$ values. In (b) and (d) also the spin-resolved signals $G_\s$ are presented. The distance between the characteristic features is shown in (a) and (d). \tr{In (a) and (e) we provide the values of the gap $\Delta_\bQ$ (they are identical in (a)-(c) and (d)-(f)).}
In the SDM case for $H \gtrsim 12 \textrm{ T}$ there are no features of the spin-up signal because the junction is transparent for incoming electrons with spin $\s = \ua$ (for explanation see main text).
}
\end{figure}

In all the following figures the parallel orientation of the $\bQ$ vector with respect to the NSJ interface has been assumed. In Fig.~ \ref{fig6} the NSJ conductance for the $s$-wave gap symmetry has been presented. Again, at high magnetic fields $H \gtrsim 12 \textrm{ T}$ the junction is transparent to incoming quasiparticles with $\s = \ua$. In the present case it is difficult to discern characteristic features of the conductance from the spin-up and spin-down channels in such a manner, that the splitting of the peaks could be measured. For this purpose, the spin-resolved signals $G_\s$ would have to be singled out, as shown in Fig.~\ref{fig6}bd, \tr{because the spin-specific features of the total conductance are subtle and could be smeared out at finite temperature or due to other effects (e.g. inelastic scattering).} Again, the characteristic features of spin-up and spin-down signals are separated by a distance equal to twice the Zeeman energy for SIM (Fig.~\ref{fig6}a) and are closer for SDM (Fig.~\ref{fig6}d).

\subsection{$d$-wave pairing symmetry}

\begin{figure*}
\includegraphics[width=17cm]{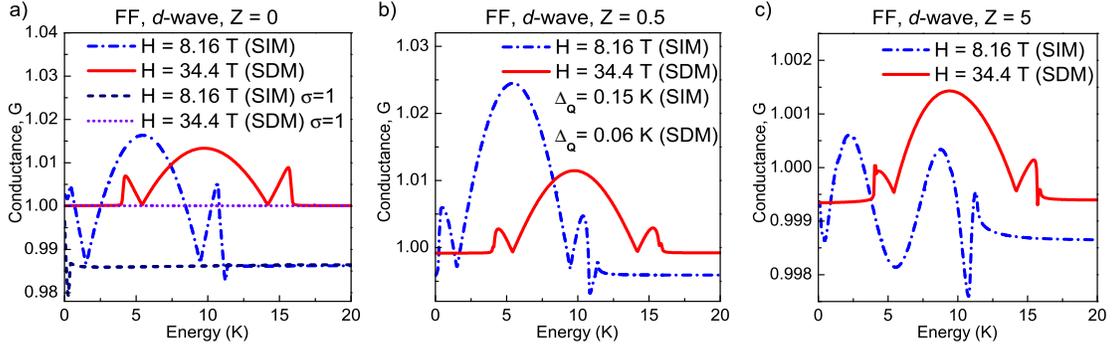}
\caption{\label{fig7} (Colour online).
Conductance spectra for the case of $d$-wave FF state for selected $Z$ values, for the SIM and SDM cases. The Cooper pair momentum is oriented along the \trr{maximum-gap (antinodal) direction} (i.e. $\theta_\bQ = 0$ and $\alpha = 0$; cf. Figs.~\ref{fig2}a and \ref{fig3}b). \tr{In (b) we provide the values of the gap $\Delta_\bQ$, and in (a) we plot also the spin-up conductance.} The magnetic field is close to $H_{c2}$. There is no clear distinct feature, which differentiates between the SIM and the SDM situations for this configuration.
}
\end{figure*}

In Fig.~\ref{fig7} the conductance in the case of FF state with $\theta_\bQ = 0$ is presented. Such phase is stable in the high-field regime (see Figs.~\ref{fig1} and \ref{fig2}). Note that by fixing the direction of $\bQ$ with respect to the NSJ interface we fix also the angle $\alpha$ (see Fig.~\ref{fig3}), as $\theta_\bQ$ is determined from the results presented in Sec. \ref{sec:sc}. Namely, the parallel vector $\bQ$ orientation with respect to the junction interface implies that $\alpha = 0$ \tr{for $\theta_\bQ = 0$ \trr{(cf. Fig. \ref{fig3}b)} and $\alpha = \pi/4$ for $\theta_\bQ = \pi/4$ \trr{(cf. Fig. \ref{fig3}c)}. In the case with $\theta_\bQ = 0$} no remarkable, model-independent differences between the SDM and the SIM cases \tr{appear, as all peaks present in Fig. \ref{fig7} come from $\sigma=\da$ electrons (see Fig. \ref{fig7}a, where the $\s=\ua$ signal has been plotted).}

\begin{figure}
\includegraphics[width=13cm]{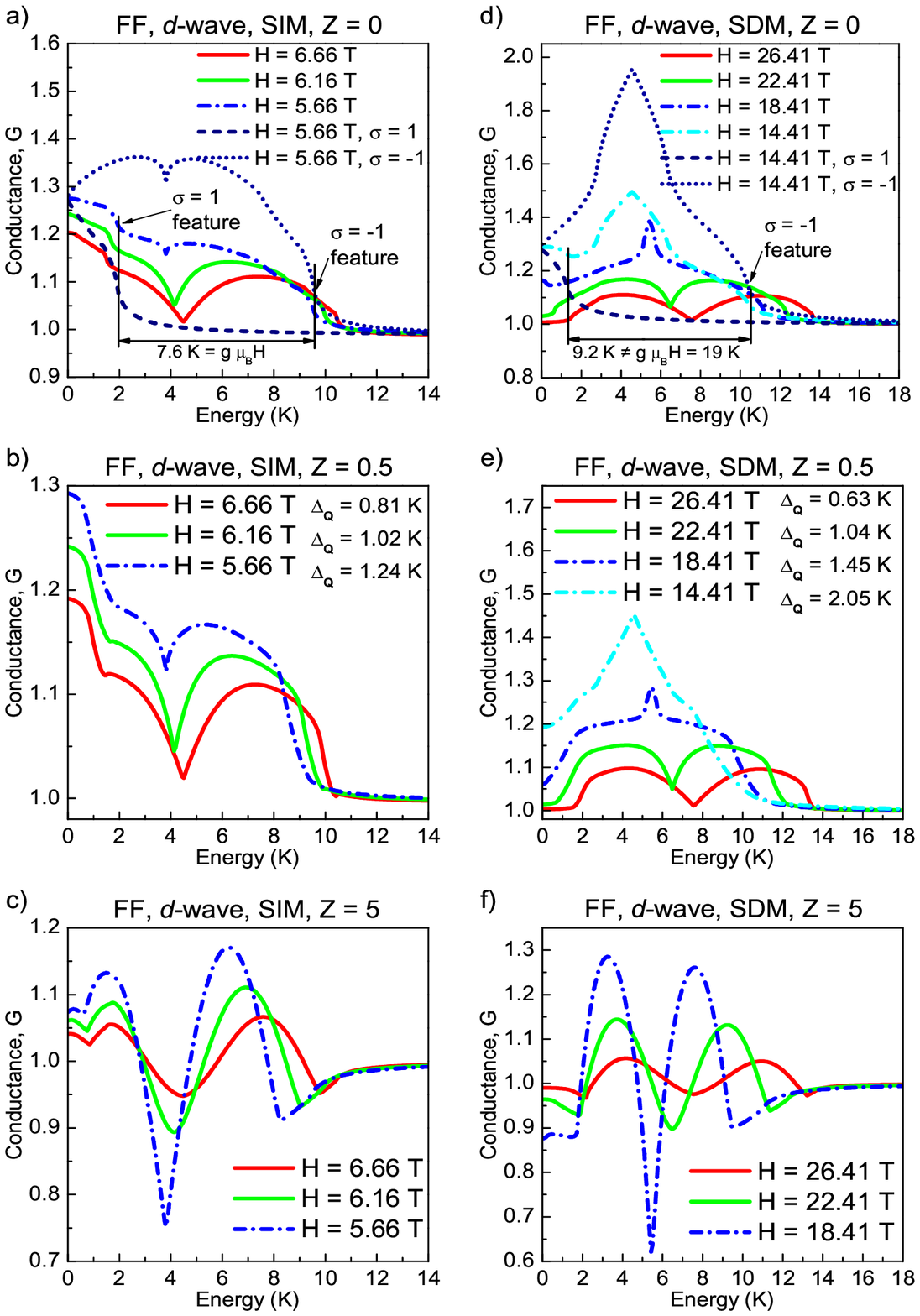}
\caption{\label{fig8} (Colour online).
Conductance spectra for the $d$-wave FF state with $\theta_\bQ = \pi / 4$ ($\bQ$ along the \trr{nodal} direction, $\alpha = \pi/4$; cf. Figs.~ \ref{fig2}a and~\ref{fig3}c) for selected barrier strengths for the SIM (a-c) and the SDM (d-f) cases. \tr{In (b) and (e) we provide the values of the gap $\Delta_\bQ$.} In (a) and (d) also the spin-resolved conductance $G_\s$ has been presented to identify spectra features for both spin channels. These features are separated by twice the Zeeman energy for the SIM, and are closer again for the SDM case.}
\end{figure}

The conductance spectra for the $d$-wave FF phase with $\theta_\bQ = \pi/4$ (with $\alpha = \pi/4$) have been presented in Fig.~\ref{fig8}. As in the $s$-wave case, and for the same reasons, at high magnetic fields the junction is transparent to spin-up quasiparticles in the SDM case. Only at $H~\lesssim~14.4~\textrm{ T}$ we were able to discern characteristic, \tr{spin-specific} features of the spectra (see Fig.~\ref{fig8}ad for the spin-resolved spectra). These features are again split by twice the Zeeman energy for SIM, and are closer for SDM. \tr{To identify the spin-specific features, spin-resolved spectra have to be analyzed, similarly as in the $s$-wave case.}

\begin{figure}
\includegraphics[width=13cm]{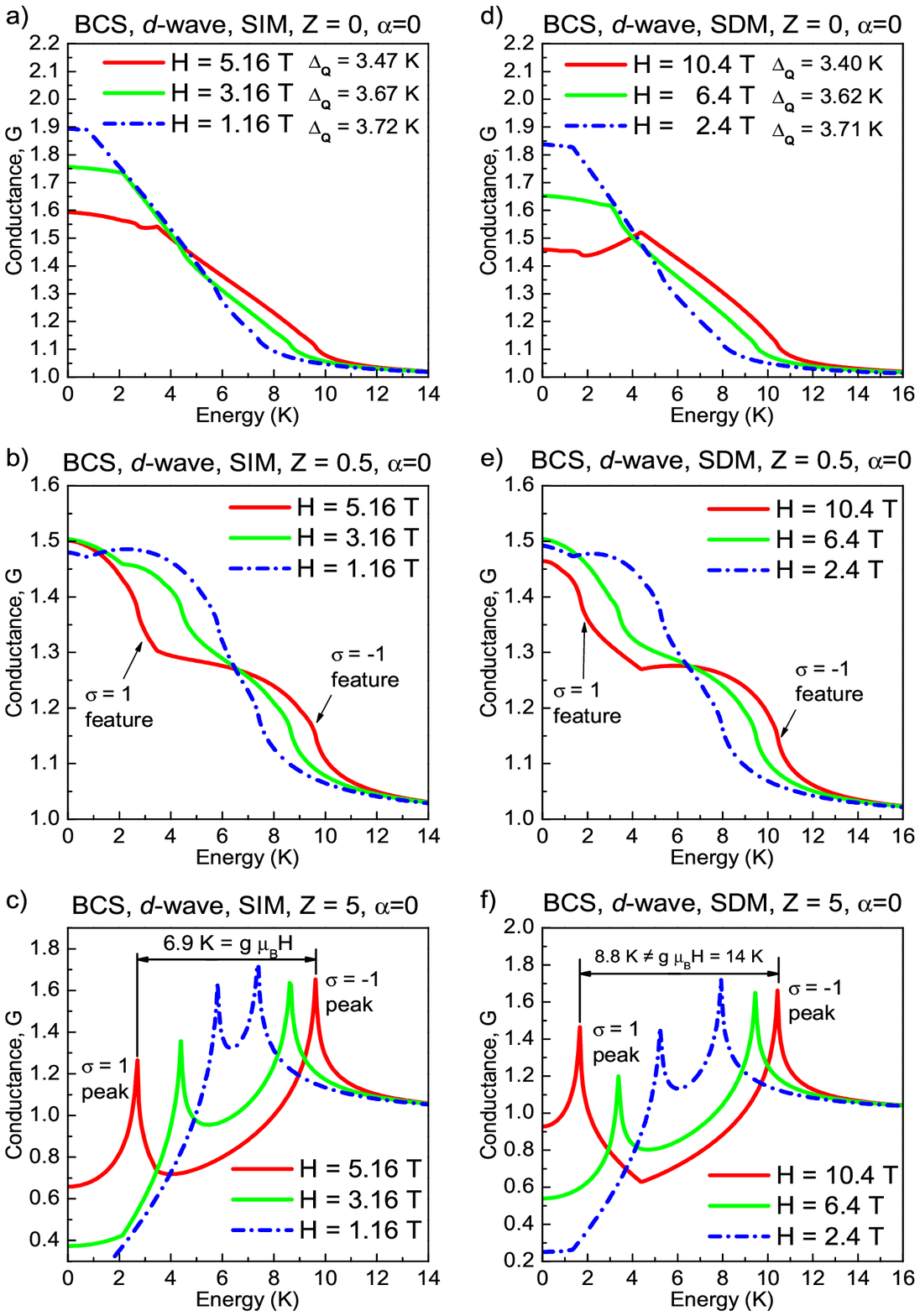}
\caption{\label{fig9} (Colour online).
Conductance spectra for the $d$-wave BCS state with ($100$) contact ($\alpha = 0$) for selected barrier strengths for the SIM (a-c) and the SDM (d-f) cases. \tr{In (a) and (d) we provide the values of the gap $\Delta_\bQ$.} In (c) and (f) in the tunneling regime ($Z=5$) conductance peaks from spin-up and spin-down channels are clearly visible already in the total conductance G. These peaks are separated by twice the Zeeman energy for SIM (c), and are closer for SDM (f).}
\end{figure}

Finally, in Fig.~\ref{fig9} we show the conductance spectra for the $d$-wave BCS state with ($100$) contact. In this case, in the tunneling limit ($Z=5$) the peaks originating from AR of quasiparticles with different spins, are most clearly visible. As previously, these peaks are split by twice the Zeeman energy for SIM, and are closer for SDM. We identify this case as the most promising for experimental verification, as discussed in the following.

\section{Relation to experiment} \label{sec:exp}

Our results imply that the splitting between the spin-up and the spin-down features of the conductance spectra is equal to twice the Zeeman energy only in the non-correlated case (SIM). In the strongly-correlated case, due to the presence of spin-dependent masses (SDM) $m_\s$ and correlation field $h_{cor}$, the separation of the spin-up and the spin-down features differs essentially. In the present case of a two-dimensional, correlated electron gas, this separation is smaller (because $m_\s$ and $h_{cor}$ compensate the Zeeman term; typically $h_{cor} \approx 0.5 \times (-h)$, cf. Refs.~\onlinecite{JKJS} and \onlinecite{JKJS2}), but in general it may be larger. For example, in the two-dimensional Hubbard model, our recent calculations \cite{SGA} yield typically $h_{cor} \approx 5 \times h$, and therefore in that model correlations \textit{enhance} splitting of the conductance peaks.

It should be in principle possible to measure the conductance-peaks splitting experimentally.
Especially, the BCS case with ($100$) contact and high barrier strength $Z$ (Fig.~\ref{fig9}cf) looks promising, as the peaks are clearly visible, and the BCS state exists in lower magnetic fields than FFLO, what should make the whole analysis simpler (the orbital effects, \cite{Benistant,*Takagaki,*Asano,*Hoppe} which may be essential especially on the normal-metal side, are less important in that regime).

Another feature differentiating the SIM case from the SDM situation is the absence of the spin-up features of conductance spectra for high magnetic fields and for the FF state. It is difficult to say, if this feature is model-independent or characteristic to the model with dispersion relation of a free-electron gas with renormalized masses.

Andreev reflection spectroscopy in magnetic field has already been reported in a few compounds. \cite{Moore,Kacmarcik,Park2,Park3,Dmitriev1,*Dmitriev2} For example in Mo$_3$Sb$_7$ point contact AR spectroscopy lead to identification of this compound as an unconventional superconductor. \cite{Dmitriev1,*Dmitriev2} Such measurements have also been performed on pure and Cd-doped CeCoIn$_5$. \cite{Park2,Park3} This compound, as a heavy-fermion superconductor and possibly host to the FFLO phase, is a natural candidate for verification of the present results. Spectra presented in Fig.~4 of Ref.~\onlinecite{Park2} resemble our Fig.~\ref{fig9}e, with splitting between the spin-up and the spin-down features of the order of $8 \textrm{ T}$ in fields of approximately $2 \textrm{ T}$. This might indicate that $h_{cor} \upuparrows h$ ($h_{cor}$  enhances $h$), but in CeCoIn$_5$ the one-band model assumed in our calculations may not be sufficient  \cite{Fogelstrom} and therefore, our interpretation is only a speculation. \footnote{Additionally, in aforementioned Fig.~4 of Ref.~\onlinecite{Park2} also antiferromagnetism plays a role in the presented spectra.} On the other hand, for a two-band model with strong correlations the $h_{cor}$ terms are also present (for both bands), and our conclusions should also hold.

Let us note that, in view of the present results, the AR spectra for the case of the BCS state with ($100$) contact and in the tunneling limit (high $Z$) would be most helpful in detecting the effect of strong correlations in superconductors. \tr{Such configuration can be studied by both Andreev reflection spectroscopy of a planar junction, as well as by the Scanning Tunneling Spectroscopy technique.}

\section{Conclusions} \label{sec:summary}

In this paper we have provided a detailed analysis of the conductance spectra of a normal metal - strongly-correlated superconductor junction.
The splitting of conductance peaks in the strongly correlated case differs from that in the uncorrelated case. It is equal to twice the Zeeman energy only in the latter case and in the correlated case it may be smaller or larger depending on the details of the electronic structure. We identify this feature as one of the {\it hallmarks of strong correlations in the superconducting phase}, as it should hold true for other models with different dispersion relations. It is most clearly visible in the case of BCS superconductor with ($100$) contact and in the tunneling regime (high $Z$). In other cases it is also present, but the spin-resolved conductances must be analyzed in order to identify the splitting unambiguously.

It would be interesting to examine other spectroscopic methods, such as the Josephson tunneling in the SQUID geometry for the systems with strong correlations (and specific features resulting from them: the spin-dependent masses and the correlation field). Such analysis should be carried out separately as it may lead to a decisively distinct interference pattern in an applied magnetic field.

\section*{Acknowledgements}
The work was supported by Ministry of Science and Higher Education, Grants Nos. N N202 128736 and N N202 173735, as well as by the Foundation for Polish Science under the "TEAM" program.

\bibliography{AndreevBib}

\end{document}